\documentclass[11pt,a4paper]{article}

\usepackage{jheppub,yfonts}

\usepackage[latin1]{inputenc}
\usepackage[T1]{fontenc}
\usepackage{amsmath}
\usepackage{graphicx}
\usepackage{caption}
\usepackage{subcaption}

\graphicspath{{figures//}}

\title{A holographic calculation of the electric conductivity of the strongly coupled quark-gluon plasma near the deconfinement transition}

\author[a]{Stefano~I.~Finazzo}
\author[a]{and Jorge Noronha}
\affiliation[a]{Instituto de F\'{i}sica, Universidade de S\~{a}o Paulo, S\~{a}o Paulo, SP, Brazil}
\emailAdd{stefano@if.usp.br}
\emailAdd{noronha@if.usp.br}

\date{\today}

\abstract{The frequency dependent conductivity $\sigma(\omega)$ of the strongly coupled Quark-Gluon Plasma (QGP) is estimated using a bottom up holographic model that can adequately describe recent lattice data for QCD thermodynamics at zero chemical potential. Different choices for the coupling between the bulk gauge field and the other bulk fields that define the background (the metric and a scalar field) are used in order to fit the lattice data for the electric charge susceptibility $\chi_2^Q/T^2$. The ratio $\sigma_{DC}/T$ is found to vary near the deconfinement transition in a way that is similar to recent lattice results. This model is used to compute the charge diffusion coefficient $D$ of the strongly coupled plasma. We find that the dimensionless combination $DT$ has the same type of temperature dependence displayed by $\sigma_{DC}/T$ and, thus, charge diffusion is suppressed at low temperatures. The frequency dependent conductivity $\sigma(\omega)$ reveals some nontrivial structure for values of the temperature near the phase transition. None of these structures appear in the associated Euclidean correlator, which we also compute. Our results suggest that the conformal invariance violation near the QCD deconfinement phase transition may be seen in the Euclidean correlator through a downward shift of its value at the minimum, which gives a rough estimate of the temperature dependence of the DC conductivity in the plasma.}

\begin{document}

\maketitle

\section{Introduction}

The gauge/gravity duality \cite{Maldacena:1997re,Gubser:1998bc,Witten:1998qj} is a powerful non-perturbative tool that can be used to investigate the transport properties of strongly coupled gauge theories with large number of colors $N_c$ \cite{CasalderreySolana:2011us}. In particular, after the seminal calculation of the shear viscosity to entropy density ratio, $\eta/s$, performed in \cite{Policastro:2001yc,Buchel:2003tz,Kovtun:2004de}, a lot of effort has been put towards the determination of other transport coefficients that can be used to fully characterize the non-equilibrium dynamics of strongly coupled plasmas, such as the QGP formed in ultrarelativistic heavy ion collisions \cite{Gyulassy:2004zy}. 

While much attention has been given to the holographic calculation of transport coefficients associated with the diffusion of energy and momentum in the hydrodynamic expansion, such as $\eta$ and also the bulk viscosity \cite{Buchel:2007mf}, much less is known about transport coefficients associated with other conserved currents such as the electric conductivity $\sigma$ and the charge diffusion coefficient $D$ (in the context of heavy ion collisions). The electric conductivity, in particular, may be relevant \cite{Tuchin:2013ie,McLerran:2013hla} for the time evolution of the strong electromagnetic fields present in non-central ultrarelativistic heavy ion collisions at RHIC and the LHC \cite{Skokov:2009qp} while \cite{Hirono:2012rt} claimed that the directed flow in asymmetric heavy ion collisions may be used to estimate the value of this coefficient in the QGP. A recent lattice QCD calculation \cite{Amato:2013naa} performed using 2 + 1 dynamical flavors found that $\sigma_{DC}/T$ is enhanced near the deconfinement transition\footnote{See \cite{AMY,CaronHuot:2006te} for studies about the electric conductivity in weakly coupled plasmas.}. A similar behavior has been found using a parton-hadron non-perturbative approach \cite{Cassing:2013iz} (other recent non-perturbative calculations include \cite{Qin:2013aaa}). Given the usual difficulties encountered in computing spectral functions from Euclidean correlators determined on the lattice, further independent confirmation of such an enhancement computed using other non-perturbative approaches, such as the gauge/gravity duality, are certainly welcome.

The conductivity in strongly coupled plasmas has been studied before using holography (see, for instance, \cite{CaronHuot:2006te,Teaney:2006nc,Herzog:2007ij,Karch:2007pd,Kovtun:2008kx,Myers:2007we,Mateos:2007yp,Cherman:2009kf,Hassanain:2010fv,Hassanain:2011fn,Patino:2012py,Jahnke:2013rca}). However, in order to understand how the strong violation of conformal invariance at temperatures $T\sim 150-300$ MeV found in current lattice QCD calculations \cite{Borsanyi:2010cj} affects the electric conductivity, it is necessary to drop the assumption of a conformal plasma. While top-down string theory constructions of non-conformal plasmas are known (see Refs.\ in \cite{CasalderreySolana:2011us}), these models cannot yet describe the specific temperature dependence of the equilibrium quantities of finite temperature QCD found on the lattice. On the other hand, bottom up holographic models in 5 dimensions involving the metric and a bulk scalar field are able to adequately describe the violation of conformal invariance seen in the thermodynamical properties of QCD at vanishing chemical potentials \cite{Gursoy:2008bu,Gursoy:2008za,Gursoy:2009jd,Gubser:2008yx,Gubser:2008ny,Noronha:2009ud,Noronha:2010mt}. One should keep in mind that such phenomenological models for the strongly coupled QGP may be only useful when $T\sim 150 - 300$ MeV. For lower temperatures an effective description involving explicit hadronic degrees of freedom should be used \cite{Karsch:2003vd,Huovinen:2009yb,NoronhaHostler:2012ug,NoronhaHostler:2008ju} while at sufficiently high temperatures a weak coupling description of the QGP is more appropriate\footnote{In fact, recent calculations \cite{Andersen:2011sf,Haque:2013qta,Mogliacci:2013mca,Haque:2013sja} involving Hard Thermal Loop perturbation theory were shown to provide a good description of the high temperature QGP properties in equilibrium.} (note also that these non-conformal holographic models remain strongly coupled even in the UV, which is not the case of an asymptotically free theory such as QCD).

A few years ago it was shown in Ref.\ \cite{DeWolfe:2010he} that the effects of a nonzero baryon chemical potential can be nicely incorporated into this class of models by adding a $U(1)$ gauge field in the bulk that is dual to the conserved baryon current at the boundary\footnote{For QCD with three dynamical quark flavors, the equilibrium pressure may depend on the baryon $\mu_B$, electric charge $\mu_Q$, and strangeness $\mu_{S}$ chemical potentials besides the temperature $T$, i.e., $p=p(T,\mu_B,\mu_Q,\mu_S)$. The case described in \cite{DeWolfe:2010he} corresponds to setting $\mu_Q =\mu_S=0$ (i.e., all quark flavors have the same chemical potential equals $\mu_B/3$).}. This general strategy follows directly from the holographic dictionary which establishes that global symmetries at the boundary are dual to gauge symmetries in the bulk \cite{Witten:1998qj}. While it is possible to include D-branes into this type of bottom up model to describe its flavor content \cite{Jarvinen:2011qe,Alho:2012mh}, the Einstein+Scalar+Maxwell model pursued in \cite{DeWolfe:2010he} contains the minimum physics needed to study the effects of global conserved charges in a strongly coupled plasma\footnote{Similar models, usually defined in asymptotically AdS$_4$ spaces, have been used in condensed matter applications \cite{Hartnoll:2009sz,Charmousis:2010zz}. See also \cite{Cai:2012xh} for applications of the Einstein+Scalar+Maxwell model in the study of the QCD phase diagram.}. 

Moreover, this type of model provides a straightforward way to compute the transport coefficients associated with the given conserved charges when their chemical potentials vanish. In fact, in this case the on-shell gauge field in the bulk vanishes and the Maxwell action enters only in the description of the small fluctuations needed in a linear response analysis. More specifically, the metric and the scalar field define the non-conformal background (taken at zero chemical potential) while the Maxwell action acts as a probe, entering only in the calculation of 2-point functions of the given channel evaluated on this background. Therefore, while the gauge field does not backreact on the background, it determines the calculation of susceptibilities and other transport coefficients such as the electric conductivity. 

In this paper, we shall use this Einstein+Scalar+Maxwell model to compute the frequency dependent electric conductivity and the charge diffusion coefficient in a strongly coupled plasma with thermodynamic properties similar to those displayed by QCD with three dynamical flavors \cite{Borsanyi:2010cj} at zero chemical potential. The non-conformal background described by the Einstein-Scalar sector a few parameters that enter in the scalar potential and are fixed to match lattice QCD thermodynamics \cite{Borsanyi:2010cj} at zero chemical potential. The gauge field couples with the metric in the usual way through the Maxwell action but it also couples to the background scalar field $\phi$. This coupling is described by an a priori unknown scalar function, $f(\phi)$, which does not affect the system's pressure though it enters directly in the calculation of the electric charge susceptibility $\chi_2^Q(T)$, as we will show below. Thus, $f(\phi)$ can be fixed by imposing that the electric charge susceptibility of the model matches the corresponding lattice data for $\chi_2^Q(T)/T^2$ \cite{Borsanyi:2011sw}. Once $f(\phi)$ is determined, one can use the holographic dictionary \cite{Son:2002sd} and extract the retarded Green's function of the electric current, which is used to compute the frequency dependent susceptibility $\sigma(\omega)$. The DC conductivity is simply $\sigma_{DC}=\lim_{\omega\to 0}\sigma(\omega)$ and it may be computed directly using the membrane paradigm \cite{Iqbal:2008by}. The charge diffusion coefficient $D$ can be directly obtained using the Einstein relation involving the $\sigma_{DC}$ and $\chi_2^Q$, which is valid for this class of theories \cite{Iqbal:2008by}. Since all the parameters of the model are fixed to match known equilibrium quantities computed on the lattice, the transport properties obtained in the model can be interpreted as holographic predictions that may be compared with the results of other methods.

This paper is organized as follows. In Section \ref{sec:model} we present the details about the holographic model used in this work. Section \ref{sec:quarksusc} is reserved to the calculation of the electric charge susceptibility $\chi_2^Q$ and its comparison to lattice data. In Section \ref{sec:cond}, we present the study of the frequency dependence of the conductivity and also compute the charge diffusion constant. The spectral function that enters in the calculation of $\sigma(\omega)$ is then used in Section \ref{sec:euc} to compute the Euclidean correlator. In Section \ref{sec:concl} we present our conclusions and outlook.

\section{Non-conformal holographic model}
\label{sec:model}

The holographic model that defines the strongly coupled plasma studied in this paper is given by the 5-dimensional action
\begin{equation}
S_{ES}=\frac{1}{2\kappa^2}\int d^5x \sqrt{-g}\left(\mathcal{R}-\frac{(\partial \phi)^2}{2}-V(\phi)\right)\,.
\end{equation}
The Ansatz for the metric used here (also known as Gubser gauge \cite{Gubser:2008ny}) is
\begin{equation}
\label{eq:gubsergauge}
ds^2=e^{2A(\phi)}\left(-h(\phi)dt^2+d\mathbf{x}^2\right)+e^{2B(\phi)}\frac{d\phi^2}{h(\phi)}
\end{equation}
where the scalar field $\phi$ is set as the fifth coordinate and $h(\phi)$ has a simple zero at the horizon $\phi=\phi_h$ while at the boundary $\phi \to 0$ and one recovers $h(\phi \to 0)=1$. The background (which describes the equilibrium properties of the plasma at nonzero $T$ and $\mu_Q=0$) is defined by the metric $g_{MN}(\phi)$ where $M,N=t,\mathbf{x},\phi$. The metric of the space-time that is the solution of Einstein's equations is assumed to be asymptotic $AdS_5$ with radius $L\equiv 1$ (this is reflected in the choice for the near boundary behavior of $V(\phi)$). In this bottom-up phenomenological approach, the scalar potential $V(\phi)$ is not determined directly from string theory. Rather, it is conveniently chosen to reproduce the temperature dependence of the equilibrium properties of the plasma such as its speed of sound. The entropy density is given by the area of the horizon
\begin{equation}
s=\frac{2\pi}{\kappa^2}e^{3A(\phi_h)}\,,
\end{equation}
and the Hawking temperature of the black brane is 
\begin{equation}
T=e^{A(\phi_h)-B(\phi_h)} \frac{| h'(\phi_h) |}{(4\pi)}.
\end{equation}

\begin{figure}
\centering
  \includegraphics[width=.5\linewidth]{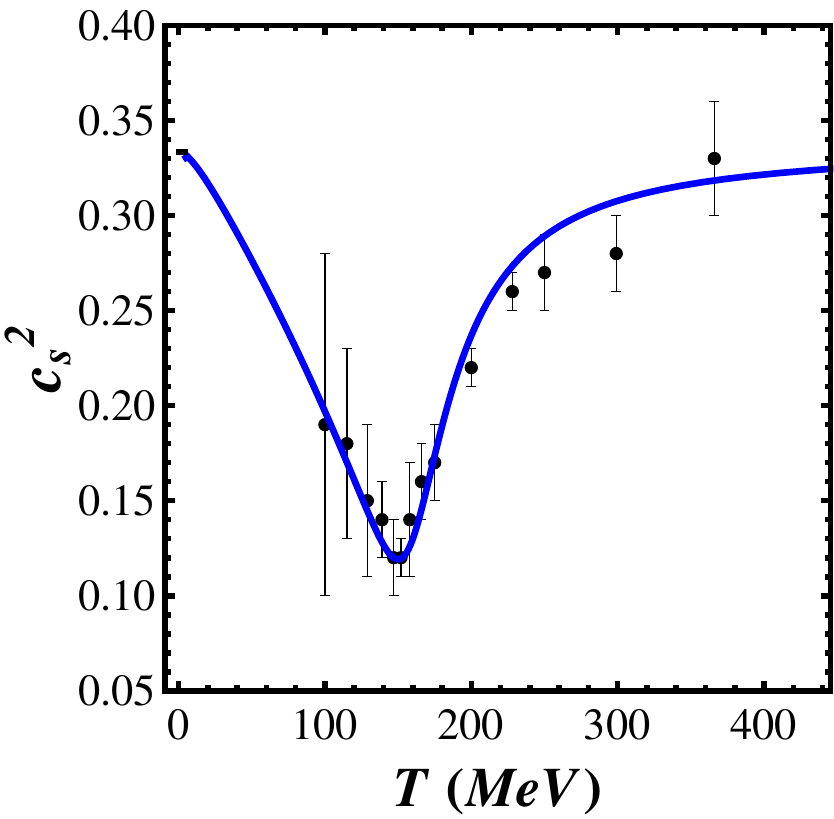}
  \caption{(Color online) The speed of sound squared of the plasma $c_s^2$ as a function of the temperature $T$ for our holographic model (solid curve) compared with lattice results for QCD with physical quark masses from \cite{Borsanyi:2010cj}.}
  \label{fig:cs2}      
\end{figure}

The numerical procedure to solve the equations of motion for the metric and the scalar field is the one derived in \cite{Gubser:2008ny} and used in the calculation of the Polyakov loop in \cite{Noronha:2009ud,Noronha:2010mt}, the heavy quark and light quark energy loss in \cite{Ficnar:2010rn,Ficnar:2011yj,Ficnar:2012yu}, and the Debye screening mass in \cite{Finazzo:topub}. A reasonable fit to the lattice data for the speed of sound squared $c_s^2 = d \log T/ d \log s$ in QCD (data from \cite{Borsanyi:2010cj} and shown in Fig.\ \ref{fig:cs2}) in the temperature interval $T \sim 150-300$ MeV is obtained using a potential similar to that studied in \cite{Gubser:2008yx},
\begin{equation}
V(\phi)=-12 \cosh \gamma \phi + b_2 \phi^2 + b_4 \phi^4 + b_6 \phi^6
\end{equation} 
where $\gamma=0.606$, $b_2=0.703$, $b_4=-0.12$, $b_6=0.0044$\footnote{Most of the needed temperature dependence of the thermodynamic quantities can be obtained using only $\gamma$, $b_2$, and (to a less extent) $b_4$. The other coefficient, $b_6$, is only needed if one wants to describe the phase transition region very accurately, as we have tried here. This choice of parameters is nearly the same as in \cite{Ficnar:2010rn}, which used older lattice data to find the set of parameters. The only difference with respect to the set used in \cite{Ficnar:2010rn} is our choice of $b_6$, which had to be updated to better describe the lattice data from \cite{Borsanyi:2010cj}. The temperature scale $T_c$ is chosen in a way that the minimum of the speed of sound squared in the model matches the value found on the lattice. This gives $T_c=150$ MeV. For more details of the thermodynamics and other properties of this model parametrization see \cite{Finazzo:topub}.}. The UV scaling dimension of the relevant operator dual to the bulk scalar field $\phi$ is $\Delta = 3.0$. Note that to have a crossover transition, as found on the lattice \cite{Borsanyi:2010cj}, the black brane solution must be the most stable solution (largest pressure) for all $T$ (in contrast to the case involving a first order phase transition studied in \cite{Gursoy:2008bu}). 

The gauge field in the background is set to zero, $A_M=0$ (remember that $\mu_Q$=0). Its fluctuations are needed to compute the retarded Green's function associated with the electric current and they are described by the action
\begin{equation}
S_{M} = -\frac{1}{2\kappa^2}\int d^5x \sqrt{-g}\,\frac{f(\phi)}{4}F_{MN}F^{MN}
\end{equation}
where the field tensor is $F_{MN}=\nabla_M A_N-\nabla_N A_M$ and $f(\phi)$ is an unknown function of the background scalar field. This function enters in the calculation of the electric charge susceptibility, $\chi_2^Q(T)=(\partial^2 p/\partial \mu_Q^2)_T$, which is defined at $\mu_Q=0$. 

\section{Electric charge susceptibility}
\label{sec:quarksusc}

We now proceed to fix the form the bulk U(1) gauge coupling $f(\phi)$. The strategy is to compute, via holography, the electric charge susceptibility and then choose a simple form for $f(\phi)$ that reproduces the corresponding lattice data for the charge susceptibility $\chi_2^Q(T)$.

In the following, we will use the membrane paradigm to compute the electric charge susceptibility. The applicability of this method for this type of calculations was discussed in detail in \cite{Iqbal:2008by} and we refer the reader to that work for the details. The most convenient gauge to work out these calculations is the conformal gauge, defined by
\begin{equation}
\label{eq:conformal}
ds^2 = e^{2\tilde{A}(z)} \left(- \tilde{h}(z) dt^2 + d\vec{x}^2 +\frac{dz^2}{{\tilde{h}(z)}}\right),
\end{equation}
where now the bulk scalar field is a function of $z$, i.e., $\phi=\phi(z)$, the horizon is at $z=z_h$ ($\phi(z \to z_h) = \phi_h$), and the asymptotically AdS$_5$ boundary is located at $z \to 0$ (where $\phi(z \to 0) \to 0$). The gauges \eqref{eq:gubsergauge} and \eqref{eq:conformal} are related by the equation
\begin{equation}
\label{eq:gaugech}
z(\phi) = \int_0^{\phi}d\phi'\, e^{B(\phi') - A(\phi')},
\end{equation}
which can be inverted to yield $\phi(z)$. Moreover, we have that $\tilde{A}(z) = A(\phi(z))$ and $\tilde{h}(z) = h(\phi(z))$. We shall use the conformal gauge in all the calculations below.

From the membrane paradigm \cite{Iqbal:2008by}, the electric charge susceptibility $\chi_2^Q(T)$ in conformal gauge is simply given by
\begin{equation}
\chi_2^Q = \frac{1}{\int_{0}^{z_h} dz\,[e^{\tilde{A}(z)} f(\phi(z))]^{-1}}\,.
\label{gubserformula}
\end{equation}
We remark that the gauge field is zero for the $\mu_Q=0$ calculations. In fact, $f(\phi)$ only enters in the calculation of $\chi_2^Q$ and $\tilde{A}(z)$, $\tilde{h}(z)$, and $\phi(z)$ are, of course, not influenced by the gauge field at $\mu_Q=0$, justifying our procedure for solving only the equations for the metric and the scalar field pursued in the previous section. We note that the proper dimensionless quantity to evaluate is $\chi_2^Q/T^2$. For a conformal field theory, $\chi_2^Q/T^2$ is a constant. For 3 flavor QCD in the Stefan-Boltzmann limit $\chi_2^{SB}/T^2= 2/3$\footnote{For convenience, we have set the electric charge to 1 in this paper.}.

Let's investigate the minimum physical requirements that the gauge coupling $f(\phi)$ must satisfy. First, $f(\phi)$ must clearly be positive and smooth in the bulk. Second, in order to recover the correct UV  fixed point behavior $\chi_2^Q/T^2 \to \, \mathrm{constant}$ for $T \to \infty$, we must require, apart from the geometry being asymptotically AdS$_5$, that $f(\phi)$ goes to a finite constant as $z_h \to 0$ ($\phi(r_h) \to 0$), in order to render the integral in \eqref{gubserformula} proportional to $T^2$. Third, in order to have $\chi_2^Q \to 0$ as $T \to 0$, we must require that $f(\phi) \to 0$ as $\phi \to \infty$ so that the integral in \eqref{gubserformula} diverges.

\begin{figure}
\centering
  \includegraphics[width=.5\linewidth]{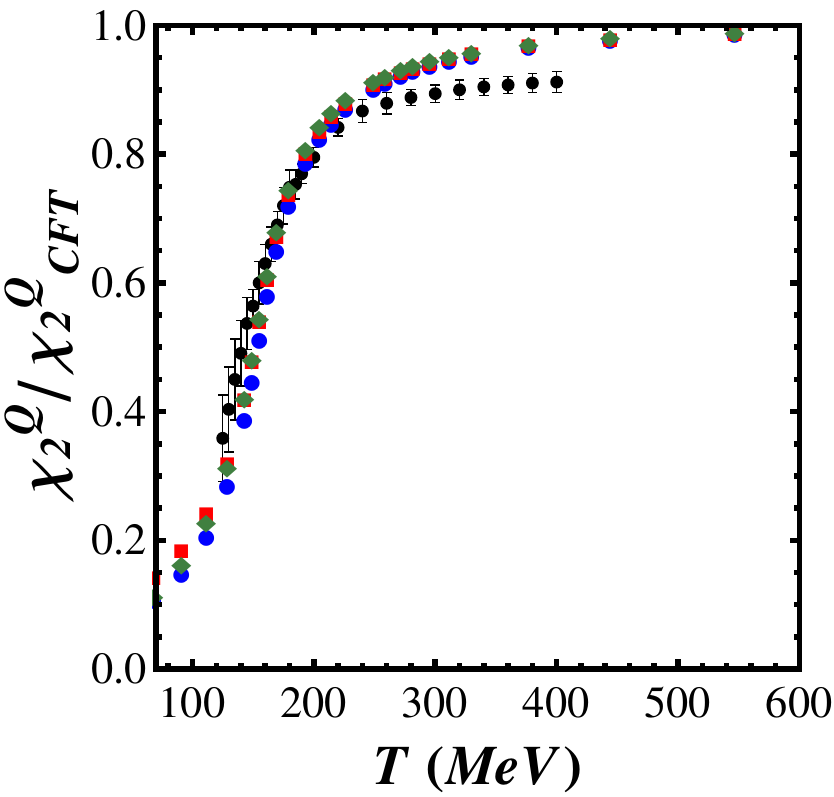}
  \caption{(Color online) The electric charge susceptibility $\chi_2^Q$ of the holographic model, normalized by its conformal limit, as a function of the temperature $T$ of the plasma. The circles, squares, and diamonds correspond to the results found using the parametrizations in \eqref{eq:fmodels}, \eqref{eq:fmodels1}, and \eqref{eq:fmodels2}, respectively. The lattice data points for $\chi_2^Q/\chi_2^{SB}$ \cite{Borsanyi:2011sw} are in black.}
  \label{fig:xi2T2}      
\end{figure}

With these requirements in mind, we have chosen three different simple parametrizations for the gauge coupling in order to check the sensitivity of the electric transport coefficients with the choice of $f(\phi)$. The parametrizations are: 
\begin{align}
\label{eq:fmodels}
f_1(\phi) = & \frac{\mathrm{sech}({a_1 \, \phi})}{g_{5,1}^2}, \\
f_2(\phi) = & \frac{1}{g_{5,2}^2}\frac{1}{(\phi^2 + a_2^2)} \label{eq:fmodels1}\quad \quad \mathrm{and} \\
f_3(\phi) = & \frac{e^{-a_3^2 \phi^2}}{g_{5,3}^2},
\label{eq:fmodels2}
\end{align}
where $a_1$, $a_2$, $a_3$, and $g_{5,i}$ are constants. In order to best fit the lattice results for $\chi_2^Q/T^2$ of Ref.\ \cite{Borsanyi:2011sw} (for another set of lattice data for $\chi_2^Q$, which are however compatible with \cite{Borsanyi:2011sw}, see Ref.\ \cite{Bazavov:2012jq}), we have chosen $a_1 = 0.4$, $a_2 = 4.0$ and $a_3 = 0.23$. We have normalized the results for $\chi_2^Q$ computed holographically using the highest temperature available numerically ($T/T_c \sim 10$) and assumed that the conformal regime $\chi_{2\,CFT}^Q$ \cite{CaronHuot:2006te} has already been reached at this temperature - this is reasonable since the holographic results reach conformality already at $T \sim 3-4 T_c$. One can see in Fig.\ \ref{fig:xi2T2} that the holographic model calculation for $\chi_2^Q/\chi_{2\,CFT}^Q$ is in good agreement with lattice results \cite{Borsanyi:2011sw} (normalized by the Stefan-Boltzmann limit) for $T < 300 \, \mathrm{MeV}$ for the three different parametrizations chosen in \eqref{eq:fmodels}-\eqref{eq:fmodels2}. For $T > 300 \, \mathrm{MeV}$ there is a sizable discrepancy. However, this is not worrisome since these holographic models are not expected to model accurately QCD at high temperatures (i.e., the weakly coupled regime).

\section{Holographic calculation of the electric conductivity and charge diffusion constant}
\label{sec:cond}

The frequency dependent conductivity associated with the conserved current operator $\hat{J}^i$ ($\mathbf{x}=x_1,x_2,x_3$) is a 3 x 3 matrix, $\sigma^{ij}(\omega)$ in Fourier space and it is directly related to the retarded Green's function of $\hat{J}^i$ via
\begin{equation}
\sigma^{ij}(\omega)=-\frac{G_R^{ij}(\omega,\mathbf{k}=0)}{i\omega}\,,
\label{conductivity1}
\end{equation}
where $G_R^{ij}(k)=-i\int d^4 x\,e^{-i k \cdot x}\theta(t)\left\langle \left[\hat{J}^i(t,\mathbf{x}),\hat{J}^j(0,\mathbf{0})    \right] \right\rangle_{T}$ (with $k_\mu=(-\omega,\mathbf{k})$). The conductivity appears in Ohm's law as $\langle \hat{J}^i(\omega)\rangle =\sigma^{ij}(\omega)F_{jt}(\omega,z\to 0)$. Rotational invariance implies that $\sigma^{ij}(\omega)=\sigma(\omega)\delta^{ij}$ and, without any loss of generality, we shall assume here that the external electric field is in the $x_1$ direction. 

\subsection{DC conductivity}

The DC electric conductivity is simply the limit $\sigma_{DC}=\lim_{\omega\to 0}\sigma(\omega)$. For the type of theory we consider in this paper, $\sigma_{DC}$ can be straightforwardly computed using the general formula derived in Eq.\ (47) of Ref.\ \cite{Iqbal:2008by} via the membrane paradigm, which gives (in conformal gauge)
\begin{equation}
\sigma_{DC} = f(\phi(z_h))e^{\tilde{A}(z_h)}.
\end{equation}
It is now clear that if $f(\phi)$ satisfies the properties given in the foregoing section, then $\sigma_{DC}/T$ goes to a constant when $T \to \infty$ (the expected conformal behavior found in \cite{CaronHuot:2006te}) and $\sigma_{DC}/T \to 0$ as $T \to 0$. Since $V(\phi)$ is completely fixed by the thermodynamics and $f(\phi)$ was fixed to reproduce the lattice data for the electric charge susceptibility, we have no more free parameters left to determine and $\sigma(\omega)$ can then be considered a prediction of the holographic model.

\begin{figure}
\centering
  \includegraphics[width=.5\linewidth]{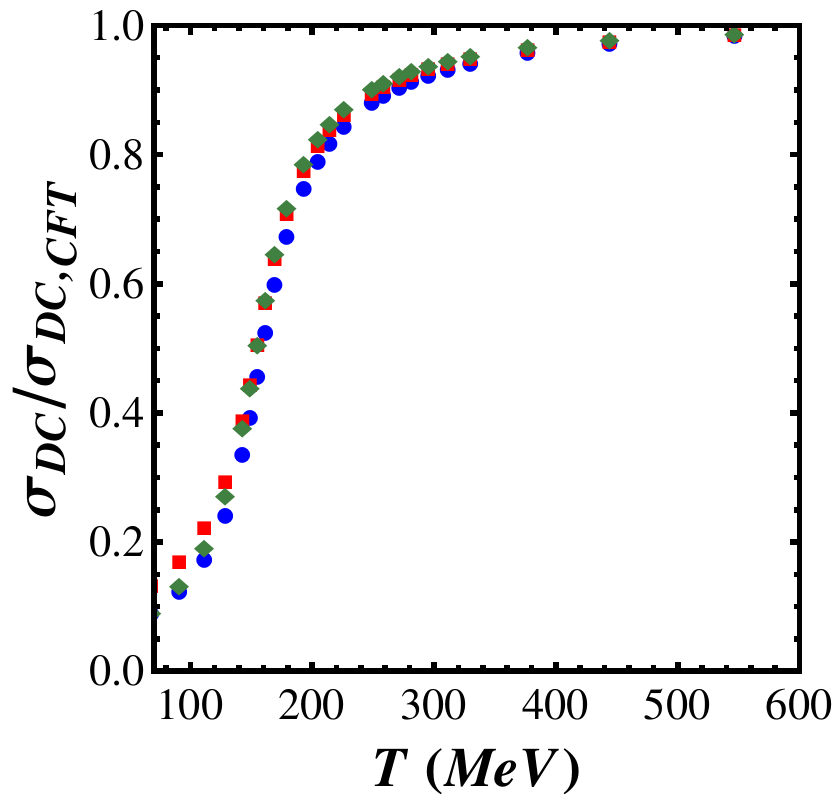}
  \caption{(Color online) The DC conductivity divided by its conformal value as a function of the temperature $T$ of the plasma. The circles, squares, and diamonds correspond to the results found using the parametrizations in \eqref{eq:fmodels}, \eqref{eq:fmodels1}, and \eqref{eq:fmodels2}, respectively.}
  \label{fig:sigmaT}      
\end{figure}

Using the parametrizations for $f(\phi)$ discussed above, we obtain the result shown in Fig.\ \ref{fig:sigmaT} for $\sigma_{DC}$, where we again normalize by the conformal result. One can see that the DC conductivity varies rapidly in the crossover region, a feature also seen in recent lattice QCD calculations \cite{Amato:2013naa}. Note also that the results for $\sigma_{DC}/\sigma_{DC, CFT}$ are robust with respect to the specific form of the gauge coupling $f(\phi)$ (though note that ours choices for this function guarantee that the conformal limit is reached from below). Also, we remark that since our charge susceptibility in principle includes the strange quark contribution, our results may be taken as estimates for the DC conductivity in the QCD plasma near the deconfinement transition (in the case of QCD the result would be then normalized by its value in the Stefan-Boltzmann limit). 

\subsection{Charge diffusion coefficient}

The small charge disturbance created by the external electric field eventually diffuses back into thermal equilibrium and this diffusion process is controlled (to lowest order in a derivative expansion) by a single transport coefficient $D$ called the charge diffusion constant. This coefficient defines the hydrodynamic mode of the $G_{R}^{x_1x_1}$ correlator \cite{forster}, which has been previously investigated in holography (see, for instance, \cite{Policastro:2002se,Kovtun:2003wp,Starinets:2008fb}). 

\begin{figure}
\centering
  \includegraphics[width=.5\linewidth]{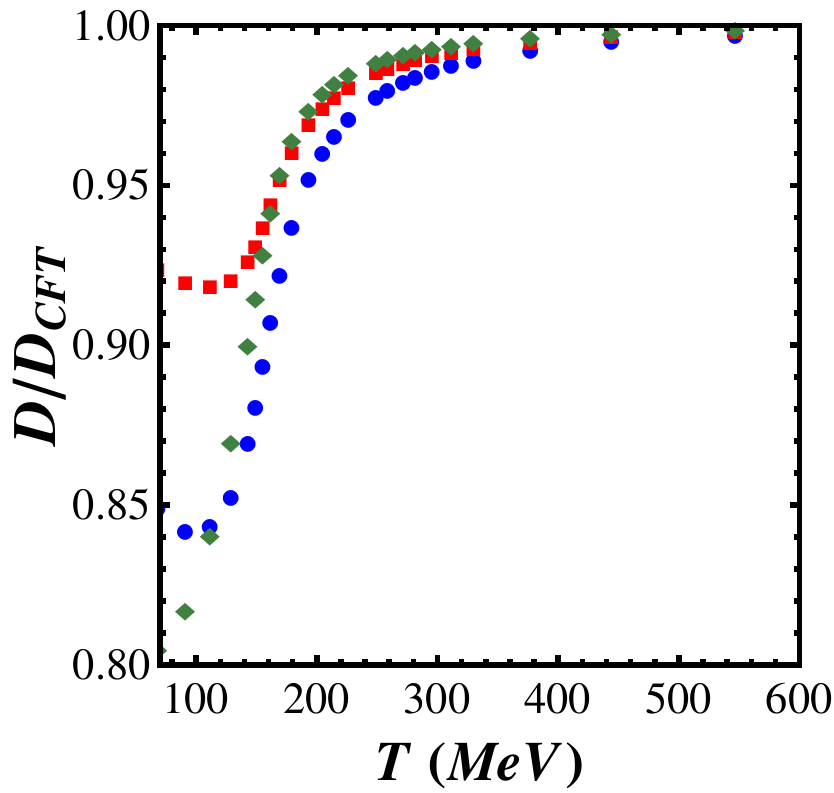}
  \caption{(Color online) The charge diffusion constant of the plasma normalized by the conformal result as a function of the temperature $T$ of the plasma. The circles, squares, and diamonds correspond to the results found using the parametrizations in \eqref{eq:fmodels}, \eqref{eq:fmodels1}, and \eqref{eq:fmodels2}, respectively.}
  \label{fig:DT}      
\end{figure}

Within the membrane paradigm, Einstein's relation among the transport coefficients involved is valid \cite{Iqbal:2008by} and the charge diffusion constant can be directly obtained using our previous results for $\chi_2^Q$ and $\sigma_{DC}$ as follows 
\begin{equation}
\label{eq:einstein}
D = \frac{\sigma_{DC}}{\chi_2^Q}\,.
\end{equation}
Thus, we may compute directly this diffusion coefficient in the dimensionless form $D/D_{CFT}$, arriving at the results shown in Fig.\ \ref{fig:DT}. Again, the results are not sensitive to the specific form of $f(\phi)$ at high temperatures $T > 150$ MeV. However, for $T < 150$ MeV, $D/D_{CFT}$ becomes very sensitive to the choice of $f(\phi)$. This does not constitute a problem per se since this holographic model certainly does not provide a good guide for the physics of the plasma at those low temperatures since the plasma is then in the hadron gas phase. However, the fact that $D/D_{CFT}<1$ at low temperatures should be robust (for instance, this behavior has been seen in the non-conformal top-down model studied in \cite{Myers:2007we}). Thus, the overall shape of the curve shown in \eqref{fig:DT} provides an estimate for the temperature dependence of the charge diffusion constant in the strongly coupled QGP, which may be checked by lattice calculations in the near future. \\

\subsection{AC electric conductivity}
\label{sec:acsigma}

To obtain the AC conductivity $\sigma(\omega)$, we must compute $G_{R}^{x_1x_1}(\omega)$. The equations of motion for the bulk fields in response to the fluctuations can be written in terms of gauge invariant quantities such as the bulk conserved current and the bulk field strength. Moreover, these equations of motion can be reduced to first order differential equations with respect to the $z$ coordinate, which completely describe the flow of the fields from the black brane horizon to the boundary \cite{Iqbal:2008by}. For nonzero momentum there are two such flow equations: one for the longitudinal channel involving the $x_1$ direction and another equation for the transverse part. However, in our case where when the momentum is taken to be zero these two equations converge (as required by rotation invariance) to the following expression \cite{Iqbal:2008by}
\begin{align}
\label{eq:floweqs}
\partial_z \bar{\sigma}(\omega,z) & = i\, \omega \,\frac{\Sigma(z)}{\tilde{h}(z)} \left[ \frac{\bar{\sigma}(\omega,z)^2}{\Sigma (z)^2}  - 1 \right]  
\end{align}
where
\begin{equation}
\Sigma(z) = f(\phi(z)) \,e^{\tilde{A}(z)}.
\end{equation}
Regularity at the horizon provides the initial condition
\begin{equation}
\label{eq:bdrycond}
\bar{\sigma}(\omega,z_h) = \sigma_{DC}
\end{equation}
and the AC conductivity is obtained by following the flow from the horizon to the boundary
\begin{equation}
\label{eq:ACcond}
\sigma(\omega) = -\frac{G_R (\omega)}{i\omega} = \bar{\sigma} (\omega,z \to 0)\,.
\end{equation} 
This gives an interpretation of $\bar{\sigma}(\omega,z)$ as the AC conductivity of the corresponding $z$-slice in the bulk. In the limit of $\omega \to 0$, the flow equation is trivial: $\partial_z \bar{\sigma} = 0$. Thus, $\bar{\sigma}$ remains at its initial value set at the horizon, which is nothing but $\sigma_{DC}$. This is the basis for the formulas used in the DC calculations in the previous section. In this case, following \cite{Iqbal:2008by}, one only needs to evaluate $\bar{\sigma}$ at the horizon to determine $\sigma_{DC}$. Now, if $\omega \neq 0$, the full flow from horizon to the boundary must be considered to determine $\sigma(\omega)$. Note that the nonlinear equation in \eqref{eq:floweqs} is a Riccati equation and, thus, it can always be rewritten in terms of a linear second order differential equation. When this is done for \eqref{eq:floweqs}, one recovers the equations of motion for the bulk field $A_{x_1}(\omega,z)$. The boundary condition \eqref{eq:bdrycond} is equivalent to impose regularity at the horizon, which in turn is equivalent to the imposition of in-falling boundary conditions at the horizon - see Appendix A of \cite{Iqbal:2008by} for details. Therefore, in this case the flow from the membrane horizon gives exactly the same results as the standard prescription used in the evaluation of holographic retarded correlators \cite{Son:2002sd}.

The numerical procedure to evaluate $\sigma(\omega)$ is straightforward. With a fixed temperature (and thus a fixed background geometry), one has to integrate \eqref{eq:floweqs} for finite $\omega$. We impose that the intercept of $\sigma(\omega)$ with the $\sigma$ axis matches $\sigma_{DC}$. The units for $\omega$ are matched by imposing the correct conformal behavior for $\omega/T \gg 1$, that is, $\sigma(\omega)/T = \mathrm{C}_{AdS_5} \times \omega/T$, where $\mathrm{C}_{AdS_5}$ is a constant found by analyzing the strongly coupled conformal limit obtained using an AdS$_5$-Schwarzchild geometry with a constant $f(\phi)$.

\begin{figure}
        \centering
        \begin{subfigure}[b]{0.5\textwidth}
				\includegraphics[width=.95\linewidth]{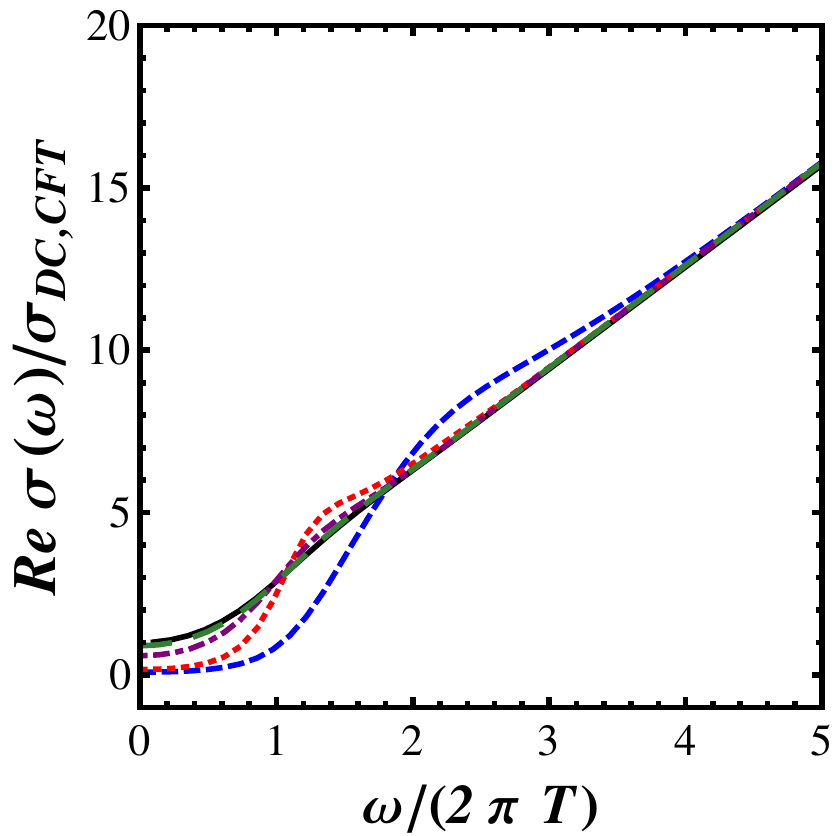}
                \caption{Model 1 - see \eqref{eq:fmodels}}
                \label{fig:sigmaACm1}
        \end{subfigure}%
        ~ 
        \begin{subfigure}[b]{0.5\textwidth}
				\includegraphics[width=.95\linewidth]{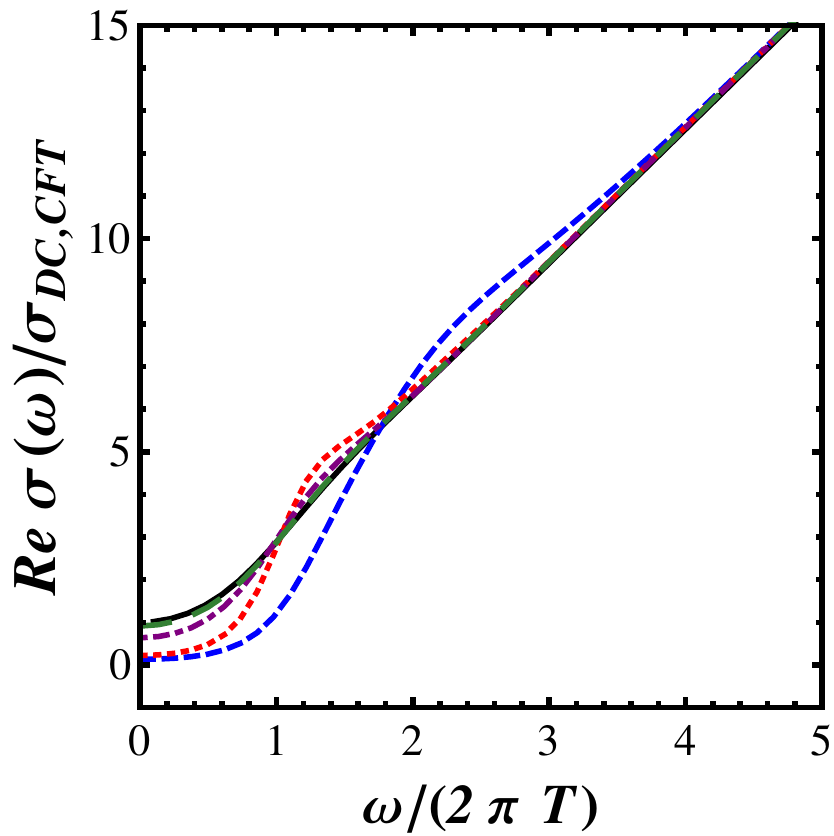}
                \caption{Model 2 - see \eqref{eq:fmodels1}}
                \label{fig:sigmaACm2}
        \end{subfigure}

        \begin{subfigure}[b]{0.5\textwidth}
				\includegraphics[width=.95\linewidth]{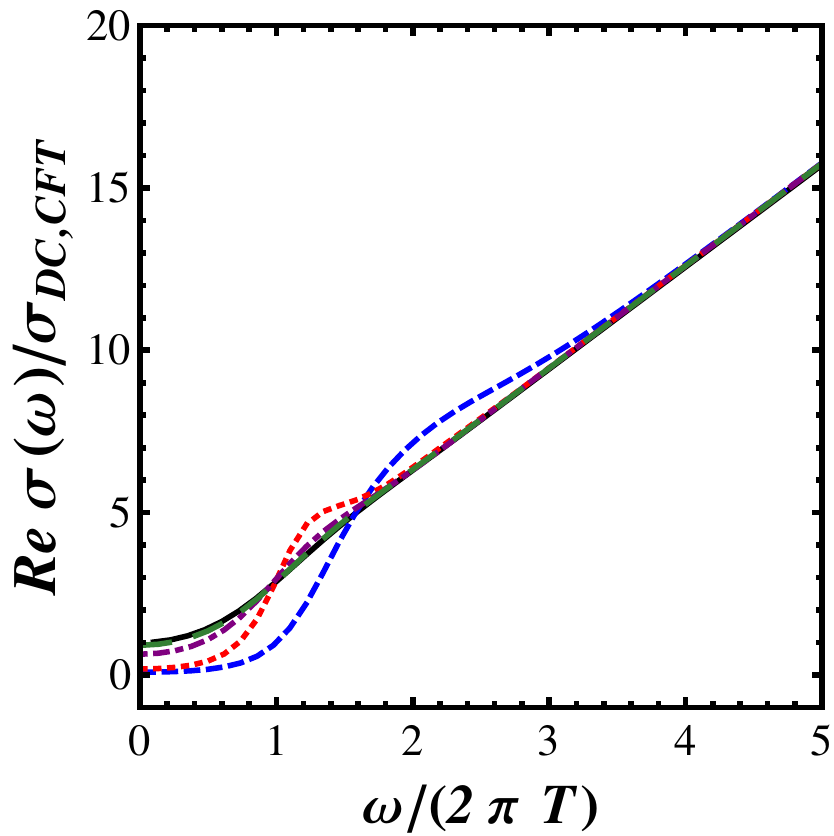}
                \caption{Model 3 - see \eqref{eq:fmodels2}}
                \label{fig:sigmaACm3}
        \end{subfigure}
		
\caption{\label{fig:sigmaAC}(Color online) The electric conductivity ${\rm Re}\,\sigma(\omega)$ (normalized by the DC conductivity in the CFT limit) as a function of $\omega/(2\pi T)$ for the different model choices of the gauge coupling in \eqref{eq:fmodels}, \eqref{eq:fmodels1}, and \eqref{eq:fmodels2}. The solid black curve is the conformal result at strong coupling, the short-dashed blue curve is for $T/T_c = 0.45$, the dotted red curve is for $T/T_c = 0.74$, the dash-dotted magenta curve is for $T/T_c = 1.13$, and the long dashed green curve is for $T/T_c = 1.81$.}
\end{figure}

\begin{figure}
        \centering
        \begin{subfigure}[b]{0.5\textwidth}
				\includegraphics[width=.95\linewidth]{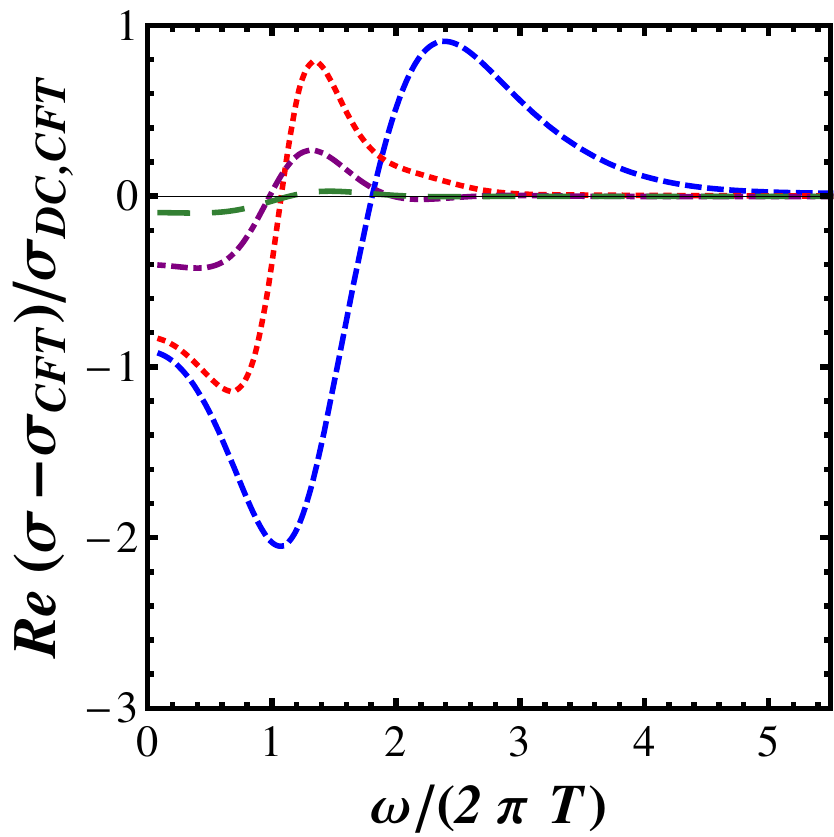}
                \caption{Model 1 - see \eqref{eq:fmodels}}
                \label{fig:sigmaACmsub1}
        \end{subfigure}%
        ~ 
        \begin{subfigure}[b]{0.5\textwidth}
				\includegraphics[width=.95\linewidth]{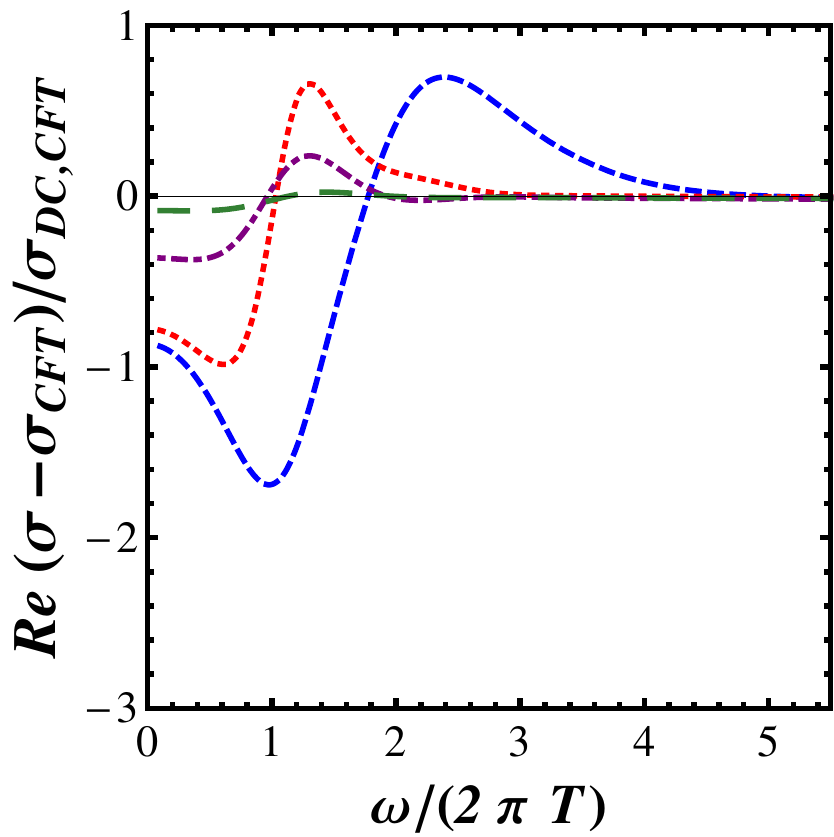}
                \caption{Model 2 - see \eqref{eq:fmodels1}}
                \label{fig:sigmaACsubm2}
        \end{subfigure}

        \begin{subfigure}[b]{0.5\textwidth}
				\includegraphics[width=.95\linewidth]{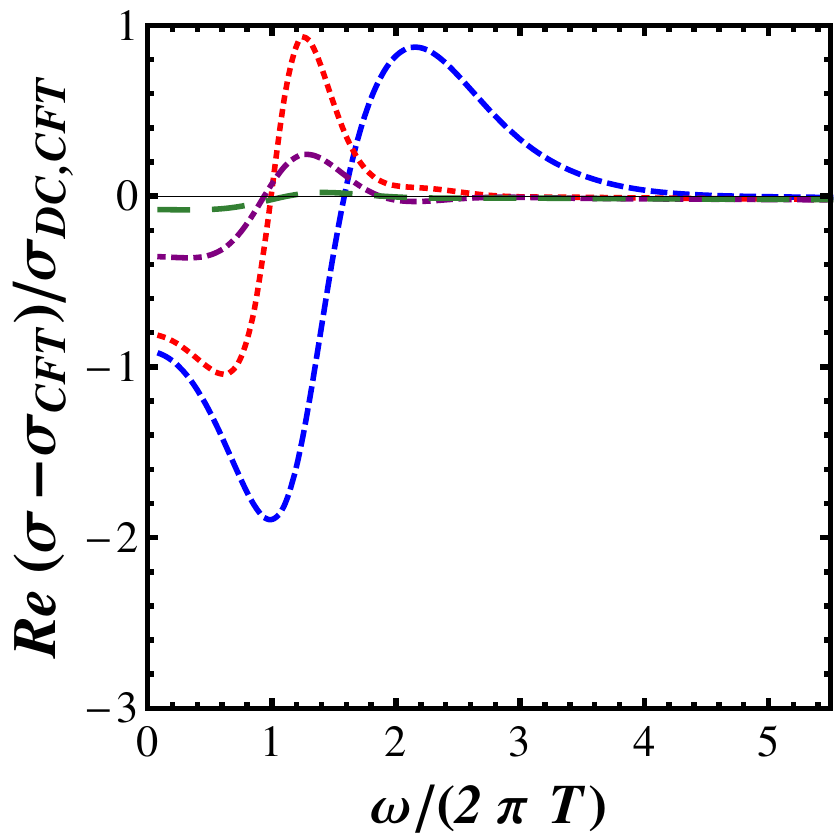}
                \caption{Model 3 - see \eqref{eq:fmodels2}}
                \label{fig:sigmaACsubm3}
        \end{subfigure}
		
\caption{\label{fig:sigmaACsub}(Color online) The electric conductivity ${\rm Re}\,\sigma(\omega)$ subtracted from the corresponding strongly coupling CFT result (normalized by the DC conductivity in the CFT limit) as a function of $\omega/(2\pi T)$ for the different model choices of the gauge coupling in \eqref{eq:fmodels}, \eqref{eq:fmodels1}, and \eqref{eq:fmodels2}. The short-dashed blue curve is for $T/T_c = 0.45$, the dotted red curve is for $T/T_c = 0.74$, the dash-dotted magenta curve is for $T/T_c = 1.13$, and the long dashed green curve is for $T/T_c = 1.81$.}
\end{figure}

Following this procedure we obtain for our three choices of $f(\phi)$ given in \eqref{eq:fmodels}, \eqref{eq:fmodels1}, and \eqref{eq:fmodels2} the results for ${\rm Re}\, \sigma(\omega)$ shown in Fig.\ \ref{fig:sigmaAC}. First, we remark that we were able to reproduce the results obtained in \cite{Teaney:2006nc} for strongly coupled $\mathcal{N} = 4$ SYM (in that case, those were interpreted as R-current correlators). We see that for $T/T_c < 1$ one can find some nontrivial structure in ${\rm Re}\,\sigma(\omega)$ when compared to the conformal strongly coupled result\footnote{We remind the reader that $T_c = 150$ MeV.}. As $T$ increases, these structures disappear. Already for $T \sim 2 T_c$, the difference between the non-conformal and conformal results is negligible. This last remark can be seen more clearly in Fig.\ \ref{fig:sigmaACsub}, where the strongly coupled conformal result has been subtracted from the non-conformal results. Also, we see that all choices for $f(\phi)$ yield similar results for the AC conductivity, in agreement with the computation of susceptibility and DC conductivity shown before. 

\section{The Euclidean correlator}
\label{sec:euc}

The AC conductivity $\sigma (\omega)$ is given by Eq.\ \eqref{conductivity1}. However, note that ${\rm Re}\,\sigma(\omega)=\rho(\omega)/\omega$, where $\rho(\omega)\equiv - {\rm Im}\, G_{R}^{x_1x_1}(\omega)$ is the spectral density. The Euclidean correlator $G_{E} (T\tau)$ in the imaginary time formalism is related to the real time spectral density by the following relation \cite{Kapusta:2006pm}
\begin{equation}
\label{eq:Euccorr}
G_{E} (T\tau) = \int_0^{\infty} d\omega \, \rho(\omega) \frac{ \cosh \left[\omega\left(T\tau - \frac{1}{2}\right)/T \right]}{\sinh \left(\omega/2T \right)}.
\end{equation}

\begin{figure}
        \centering
        \begin{subfigure}[b]{0.5\textwidth}
				\includegraphics[width=.95\linewidth]{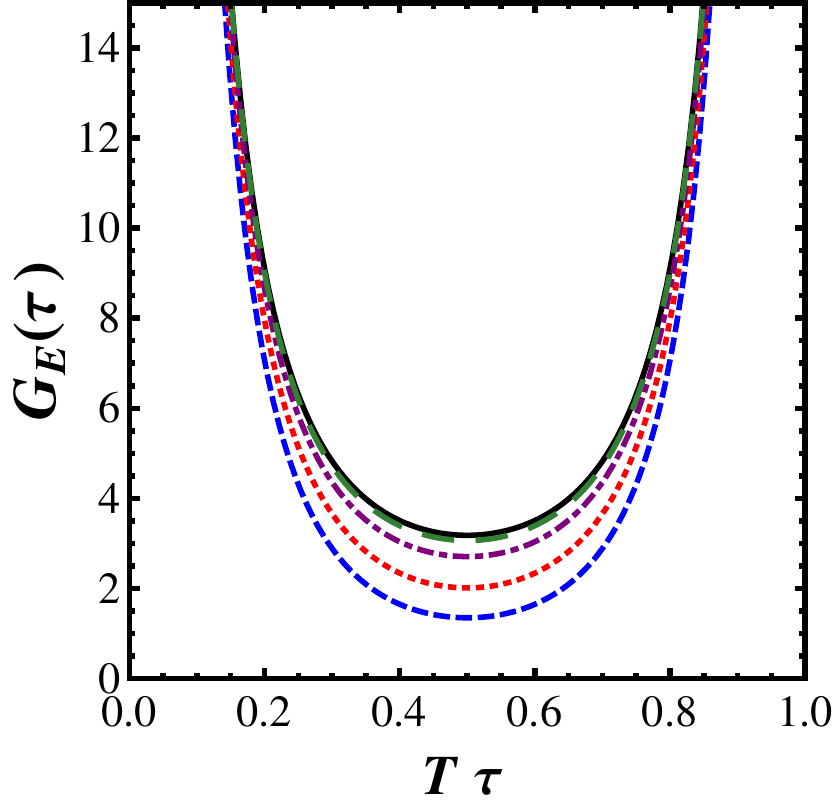}
                \caption{Model 1 - see \eqref{eq:fmodels}}
                \label{fig:GEucm1}
        \end{subfigure}%
        ~ 
        \begin{subfigure}[b]{0.5\textwidth}
				\includegraphics[width=.95\linewidth]{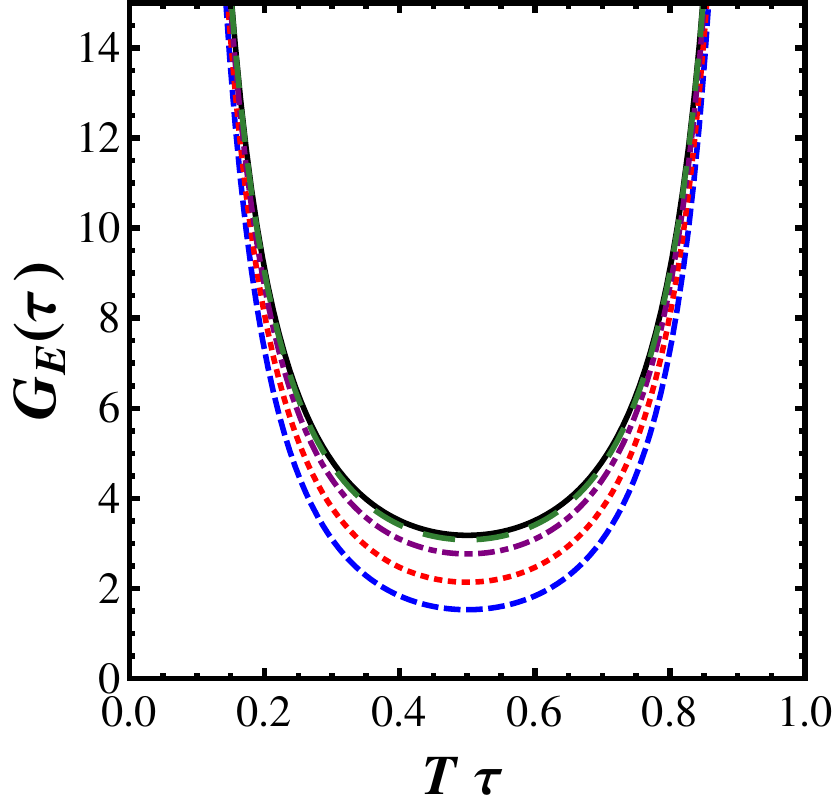}
                \caption{Model 2 - see \eqref{eq:fmodels1}}
                \label{fig:GEucm2}
        \end{subfigure}

        \begin{subfigure}[b]{0.5\textwidth}
				\includegraphics[width=.95\linewidth]{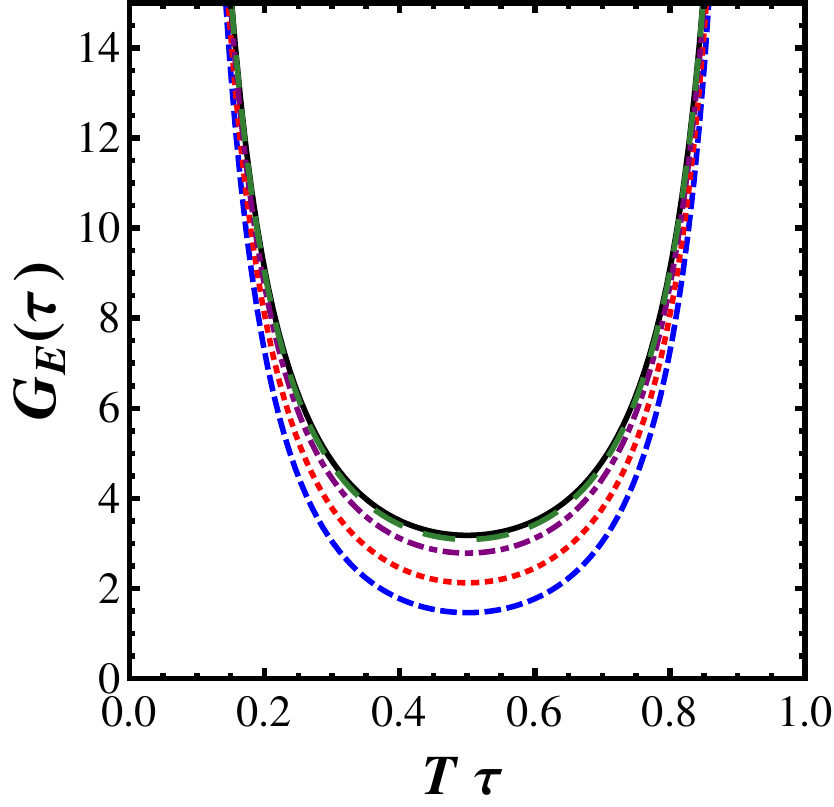}
                \caption{Model 3 - see \eqref{eq:fmodels2}}
                \label{fig:GEucm3}
        \end{subfigure}
		
\caption{\label{fig:GEuc}(Color online) The Euclidean correlator $G_{E} (T\tau)$ in \eqref{eq:Euccorr} as a function of $T \tau$. The solid black curve is the conformal result at strong coupling, the short-dashed blue curve is for $T/T_c = 0.45$, the dotted red curve is for $T/T_c = 0.74$, the dash-dotted magenta curve is for $T/T_c = 1.13$, and the long dashed green curve is for $T/T_c = 1.81$.}
\end{figure}

It is interesting to check if the structures observed in ${\rm Re}\,\sigma(\omega)$ or, alternatively, in $\rho(\omega)$ due to the strong violation of conformal invariance experienced by the theory near the deconfinement transition are reflected at all in $G_E(T\tau)$. Using Eq.\ \eqref{eq:Euccorr} with the results of the previous section, we evaluate $G_{E}(T\tau)$ for a range of temperatures and for all the three model choices of $f(\phi)$, obtaining the results shown in Fig.\ \ref{fig:GEuc}. One can see that for all the different temperatures considered that the Euclidean correlator is basically featureless - the details present in $\rho(\omega)$ are smoothed out in the computation of the Euclidean correlator. The strongly coupled CFT limit is reached already in this case at fairly intermediate temperatures, $T \sim 2 T_c$. 

Also, all model choices of the gauge coupling yield similar results - displaying the consistency already seen in the calculations done in the previous sections. This suggests that in order to obtain the real time spectral density at strong coupling $\rho(\omega)$ from $G_{E}(T\tau)$ (reversing the direction of calculation) one needs to be able to evaluate the Euclidean correlator with extremely great precision, as already remarked in the previous analysis of Teaney in \cite{Teaney:2006nc}. However, it is interesting to see that as the temperature is lowered towards the phase transition region the value of $G_{E}(T\tau)$ at the minimum (which must be at $\tau T=1/2$) decreases. This is consistent with the behavior observed in Fig.\ \ref{fig:sigmaACsub}: for lower temperatures the region in $\omega$ for which $\rho(\omega) < \rho(\omega)_{CFT}$ becomes larger and, thus, for $T\tau=1/2$ one should expect that the value of the integral performed with the conformal spectral density should be larger than the value found for the non-conformal theory. Also, this is consistent with the fact that $\sigma_{DC}/\sigma_{DC,CFT} < 1$ for those temperatures. Thus, at least within this model the downward shift of the minimum of the Euclidean correlator is a good indicator of the temperature dependence of the DC conductivity. This also seems to be the case in recent lattice calculations \cite{Amato:2013oja}.

\section{Conclusions}
\label{sec:concl}

In this paper a non-conformal, bottom-up holographic model that is able to describe recent lattice QCD thermodynamics at zero chemical potential \cite{Borsanyi:2010cj} was used to estimate the electric transport properties of the strongly coupled QGP near the deconfinement phase transition. In order to access the electric properties of the plasma, the coupling between the bulk fields that define the background (the metric and a scalar field) and the bulk gauge field (which describes the conserved current in the gauge theory) was fixed by imposing that the charge susceptibility of the model agrees with recent lattice data \cite{Borsanyi:2011sw} near the transition. All the parameters of the model were then fixed and the model was subsequently used to compute the frequency dependent electric conductivity (which has the DC conductivity as its $\omega \to 0$ limit) and the charge diffusion constant. 

The ratio $\sigma_{DC}/\sigma_{DC,CFT}$ was found to vary very rapidly in the temperature range $T \sim 150-300$ MeV, which may have some interesting implications for heavy ion collision observables \cite{Tuchin:2013ie,McLerran:2013hla,Skokov:2009qp,Hirono:2012rt}. Also, we have shown that the charge diffusion constant of the plasma has a similar temperature dependence (when normalized by its conformal value) as $\sigma_{DC}/\sigma_{DC,CFT}$ when $T>150$ MeV. Overall, we find that both the DC conductivity and the charge diffusion coefficient are suppressed with respect to their CFT values at low temperatures where the violation of conformal invariance is large\footnote{This infrared suppression was also observed in calculations performed within the soft-wall model \cite{Karch:2006pv} done in \cite{Atmaja:2008mt}.}. It would be interesting to check if that is also going to be the case in lattice calculations (in this case the high $T$ is a weakly interacting CFT). The results for ${\rm Re}\,\sigma(\omega)$ show distinct differences for temperatures below and above $T_c =150$ MeV. Below $T_c$, the violation of conformal invariance makes ${\rm Re}\,\sigma(\omega)$ smaller than its CFT value for low $\omega$ (this is consistent with our findings that $\sigma_{DC}/\sigma_{DC,CFT}\leq 1$) while it approaches the CFT result from above at high frequencies. 

We also computed the Euclidean correlator $G_{E} (T\tau)$ and its overall shape seems to be insensitive to the structure present in $\sigma(\omega)$. This means that, at least from the viewpoint of this holographic setup, the extraction of the spectral density from $G_E(T\tau)$ by analytic continuation may require very precise numerical results for the Euclidean correlator. However, within this model the downward shift of the minimum of the Euclidean correlator due to non-conformal effects seems to be a good indicator for the temperature dependence of the DC conductivity. 

A generalization of the flow equation in \eqref{eq:floweqs} can be solved numerically for nonzero momenta yielding the complete spectral density $\rho(\omega, \mathbf{k})$, which can then be used to estimate holographically the photon and dilepton production rates in the QGP near the deconfinement transition. Such study was already done for $\mathcal{N}=4$ SYM in Ref. \cite{CaronHuot:2006te} and it would be interesting to compute these observables with the model used in this paper. We intend to pursue this study in the future.

\acknowledgments

This work was supported by Funda\c c\~ao de Amparo \`a Pesquisa do Estado de
S\~ao Paulo (FAPESP) and Conselho Nacional de Desenvolvimento Cient\'ifico e
Tecnol\'ogico (CNPq). The authors thank E.~Kiritis and C.~N\'u\~nez for discussions and R.~Rougemont for comments on the manuscript.  

\appendix

\end{document}